# Wide bandwidth nanowire electromechanics on insulating substrates at room temperature

T. S. Abhilash, John P. Mathew, Shamashis Sengupta, M. R. Gokhale, Arnab Bhattacharya, and Mandar M. Deshmukh

Department of Condensed Matter Physics and Materials Science, Tata Institute of Fundamental Research, Homi Bhabha Road, Mumbai 400005, India
E-mail: deshmukh@tifr.res.in

## Abstract

We study InAs nanowire resonators fabricated on sapphire substrate with a local gate configuration. The key advantage of using an insulating sapphire substrate is that it results in a reduced parasitic capacitance thus allowing both wide bandwidth actuation and detection using a network analyzer as well as signal detection at room temperature. Both in-plane and out-of-plane vibrational modes of the nanowire can be driven and the non-linear response of the resonators studied. In addition this technique enables the study of variation of thermal strains due to heating in nanostructures.

Nanoelectromechanical systems (NEMS) have been explored in the field of sensing, and for probing nanoscale physics in recent years.[1–5] NEMS resonators of high quality factor (Q) have been demonstrated using carbon nanotubes (CNT), graphene and semiconductor nanowires (NW).[6–9] Ease and control of growth process make nanowires suitable candidates for NEMS.[8] Previous studies have reported electromechanics of InAs nanowire resonators with high mechanical responsivity.[9] Radio frequency transduction has recently been achieved with suspended graphene on $Si/SiO_2$ substrates with local gates.[10] However in these devices, the measurements are performed at low temperatures to reduce the signal-to-background ratio resulting from parasitic capacitance due to conducting Si substrates.

Insulating substrates are an alternative to minimize parasitic capacitance that arises from finite conductivity of semiconductor substrates. Sapphire substrates are suitable candidates for such devices due to reduced charge noise, and are commonly used in microwave devices and qubits;[11,12] moreover, the measurements can be carried out at room temperature (RT) without having to cool the devices to freeze the carriers. However, charging effects during electron beam lithography limit the use of such substrates for device fabrication. We have developed a simple fabrication process that overcomes charging effects while patterning on such substrates.

Transduction techniques based on mixing schemes are widely used for NEMS resonators,[13–15] however, the low measurement bandwidth of the down-mixed signal (~ kHz) is a drawback. Using our devices fabricated on sapphire we implement a fast readout scheme (~ MHz)[10,16–19] using the network analyzer that has previously been used at cryogenic temperatures. Such NEMS devices fabricated can be used for mass sensing[1,20] at room temperature and for studying strong coupling between mechanical modes with a larger bandwidth[21,22] to realize pump and probe experiments. Here we demonstrate the room temperature electromechanics of InAs nanowire resonators on insulating sapphire substrate. The mechanical motion is capacitively actuated and electrically readout using a network analyzer.[10,16–19] We show that this is possible even with smaller capacitance coupling (~ 10 aF) compared to larger capacitance (~ 40 aF) coupling used for graphene devices.[10]

InAs nanowires of length ~ 10 $\mu$m and diameter ~ 100 nm are grown[23] using metal organic chemical vapour deposition (MOCVD) via a vapor-liquid-solid process. The substrate used for fabrication of the resonators is (400 $\mu$m thick) c-plane sapphire. The device fabrication process is shown in Fig. 1. Before lithographic patterning a thin aluminum (Al) layer (20 nm) is thermally evaporated to avoid charging effects of the substrate. This also prevents breakage of nanowires during plasma etching stage of the fabrication process. The NWs are dispersed on predefined markers on the substrate and sandwiched between four layers of e-beam resists (Fig. 1a). A layer of Al (10 nm) is deposited on top to further reduce charging.

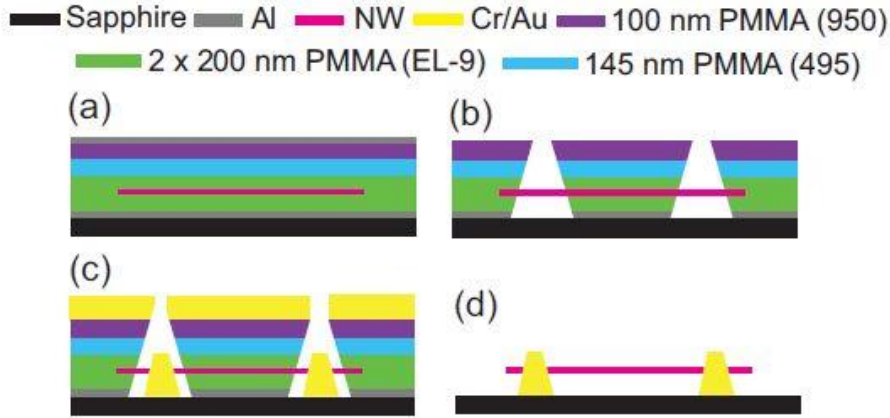

Figure 1: Steps used in the fabrication of side gated resonators on sapphire substrate. (Side gate not shown.) a) Deposition of Al layers and sandwiching the NWs between e-beam resists. b) Patterning the electrodes, development and Al layer etch. c) Deposition of chromium and gold by sputtering to fabricate source, drain and gate electrodes. d) Lift-off and removal of the Al layer.

Using electron beam lithography, source-drain electrodes and side gate are patterned. The source-drain electrodes act as mechanical clamps for the suspended NW. The spacing between the side-gate electrode and the edge of nanowire can be designed to between 50 - 300 nm. After exposure, the top Al layer is etched using photoresist developer (MF-26A). The exposed region is then developed using a mixture of methyl isobutyl ketone and isopropyl alcohol. The Al layer in the developed region is further etched using photoresist developer (Fig. 1b). Chromium (75 nm) and gold (300 nm) are sputtered to form the electrodes. To ensure good ohmic contacts, oxide / resist residues on the nanowire are removed by in-situ plasma etching before metal deposition. Lift-off is carried out in acetone to remove the polymer resist and metal layers. The remaining Al layer is etched and the substrate is rinsed in DI water. Fig. 2a shows a scanning electron microscope (SEM) image of a side-gated InAs nanowire resonator. The circuit[10] used to actuate and detect mechanical oscillations of the NW is shown in Fig. 2c.

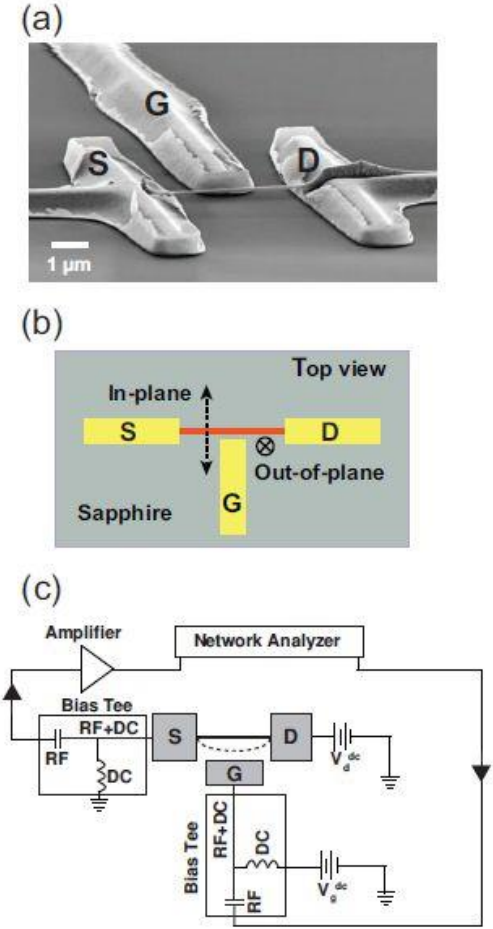

Figure 2: a) SEM image of a side gated InAs nanowire resonator on sapphire. S: source, D: drain, G: gate. b) A schematic of the in-plane and out-of-plane modes of motion of the NW. c) Schematic of the circuit used for actuation and detection of mechanical resonance of the suspended InAs nanowire resonators.

A radio frequency (rf) drive ($\tilde{V}_g$) is generated using a network analyzer (Agilent- 4396B) and a dc voltage $V_g^{dc}$ is combined using a bias-tee and applied to the side gate. A dc bias $V_d^{dc}$ is applied to the drain. The source current is split into dc and rf with a second bias-tee and the rf component is carried to a low-noise preamplifier (SRS-SR445A) providing 40 dB gain and fed into the network analyzer. The capacitive coupling between the NW and the side gate actuates the NW motion in a plane parallel to the substrate. We measure the transmission coefficient of the device $|S_{21}| = 50 \, \Omega \, \tilde{I}/\tilde{V}_g$; where 50 $\Omega$ is the network analyzer's input impedance. The current through the resonator is given by:[10]

$$\tilde{I} = j\omega C_{tot}\tilde{V}_g - j\omega \frac{\tilde{z}}{z_0}C_g V_g + V_d \frac{dG}{dV_g}\tilde{V}_g - V_d \frac{dG}{dV_g}\frac{\tilde{z}}{z_0}V_g \qquad (1)$$

where, $C_{tot}$ is the total capacitance of the device, $\omega$ is the driving frequency, $C_g$ and $z_0$ are the capacitance and distance between suspended nanowire and side gate and $\tilde{z}$ is the oscillation amplitude. The measurements are performed at room temperature (300 K) using an rf probe station in vacuum ($\sim$ 1 mbar). Prior to each measurement the network analyzer is calibrated to null the background with an applied $V_d^{dc}$ and with $V_g^{dc} = 0$ (details given in supplementary material).

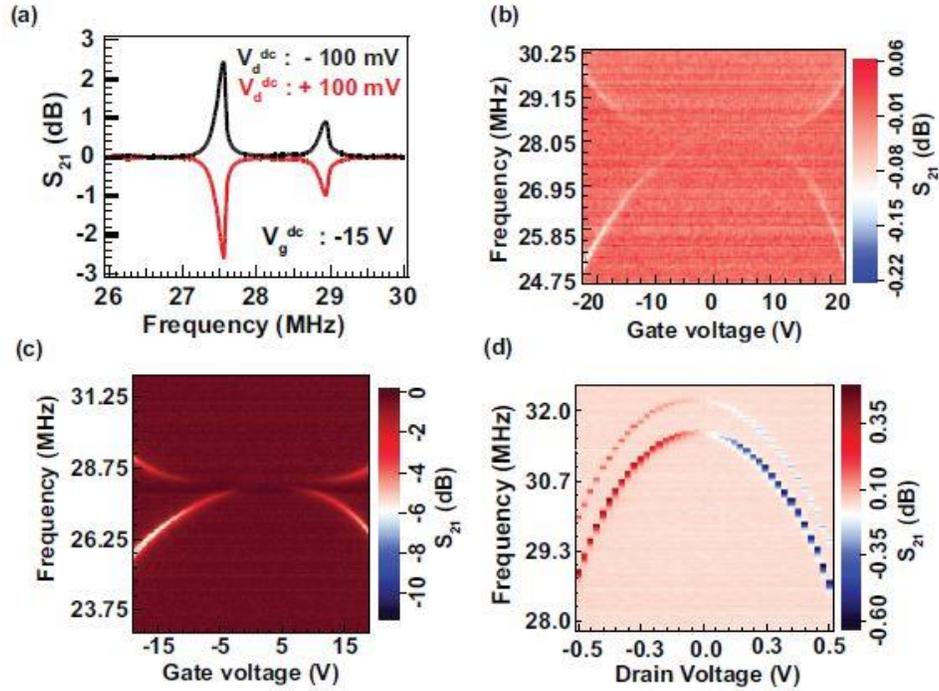

Figure 3: Measurements at RT: a) Line plots of $S_{21}$ as a function of driving frequency at $V_d^{dc} = -$ 100 mV and + 100 mV showing in-plane mode ($\sim$ 27.5 MHz) and out-of-plane mode ($\sim$ 29 MHz). b) Resonance frequency as a function of $V_g^{dc}$ at $V_d^{dc} = 0$. c) Measured resonance response as a function of $V_g^{dc}$ at a drain bias of $V_d^{dc} =$ 100 mV and d) as a function of $V_d^{dc}$ for $V_g^{dc} =$ -30 V.

Typical frequency scans of the transmission coefficient $|S_{21}|$ are shown in Fig. 3a. The resonance appears as a peak for negative $V_d^{dc}$ and a dip for positive $V_d^{dc}$. Fig. 3b and 3c shows the dispersion of resonant frequency with $V_g^{dc}$. Both in-plane and out-of-plane vibrational modes are observed. In the in-plane mode the nanowire oscillates parallel to the substrate while oscillations in the out-of-plane mode are perpendicular to the substrate (see Fig. 2b). The in-plane mode shows larger $S_{21}$ response. Resonance frequency is seen to be ~ 27.5 MHz for the in-plane mode and ~ 29 MHz for the out-of plane mode at room temperature. At low dc gate voltages, we see a feeble response from the out-of-plane mode. However, by increasing the $V_g^{dc}$, we are able to tune both the two modes into resonance. The presence of an out-of-plane degree of freedom is attributed to inhomogeneity in the side gate field.[24] The observed negative dispersion of the in-plane mode is attributed to a decrease in the effective spring constant with increasing magnitude of gate voltage.[25–27] The out-of-plane mode, however, shows a positive dispersion with gate voltage as a result of the increased tension in the NW.[25–27]

In Fig. 3b we see that both the resonance modes are observed even at zero source-drain bias. We also see that the resonance frequency is easily tuned by the source-drain bias (Fig. 3d). The calculated value of oscillation amplitude at mechanical resonance, comes out to be ~ 0.7 nm (Fig. 4 of supplementary material).

In addition to this we have explored the effect of $V_d^{dc}$ on device performance beyond the description of Eq. 1. Fig. 3d shows the dispersion in the frequency as a function of $V_d^{dc}$. Interestingly we see that the both the modes soften significantly as the bias is increased to 0.5 V. This cannot be understood on the basis of electrostatics of the device. A simple model taking into account the heating of the device and consequent reduction in the strain in the nanowire due to thermal expansion allows us to explain the reduction in the resonant frequency (the details regarding the model are explained in the supplementary information). Taking into account the thermal conductivity of the material we find that the temperature in the middle of the nanowire for a bias of 0.5 V corresponds to a temperature of 370 K in the middle of the nanowire[28] (Fig. 8 of supplementary material). The corresponding thermal expansion of the nanowire is only ~ 1 nm. However, the change in strain (~ $2 \times 10^4$) due to expansion can cause a change[29] of a few MHz in resonant frequency of the nanowire.

The resonance curve for the in-plane motion is fit to a Lorentzian line shape and the Q is extracted. The calculated quality factor of the device is 360 at RT (Fig. 4a). At low temperatures, we observe an increase in the resonant frequency (Fig. 5 of supplementary material). This is due to the increase in tension resulting from relative contraction of NW and metal electrodes. The calculated quality factor of the device is 7798 at 77 K (Fig. 6 of supplementary material) and 9958 at 5 K (Fig. 4b).

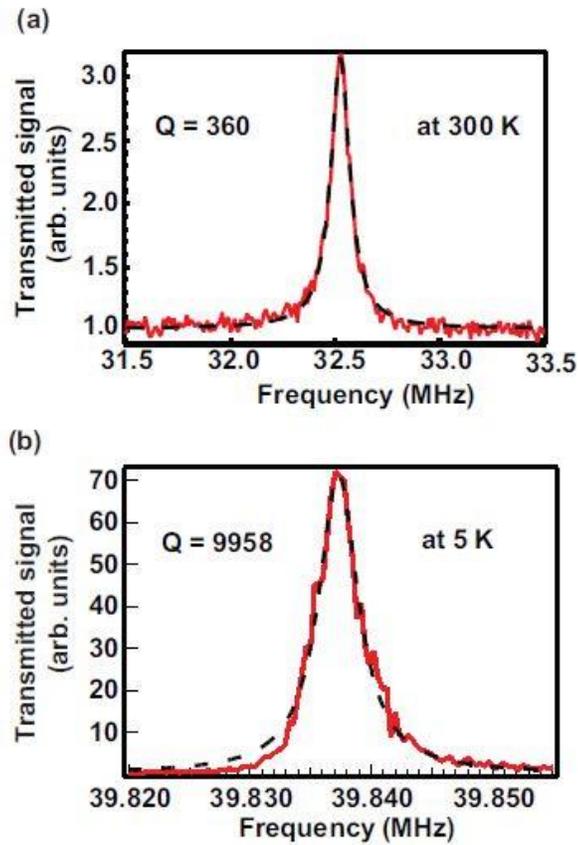

Figure 4: Quality factor of device. a) Red line shows the measured response and dotted line its Lorentzian fit of a device at RT. b) Response at 5 K. Dotted line is a fitted curve.

We have also probed the device resonance as a function of drive power. As the power is increased, the response becomes non-linear.[30–32] Fig. 5a shows the response of the wire in the linear and non-linear regimes, where the vibration is in the plane of the gate. A two dimensional plot of the resonator in the non-linear regime in Fig. 5b.

Even at high drive powers (-2 dBm) non-linearity is observed only in the in-plane mode. The ability to control the extent of non-linearity in a mode is of interest as it opens up the possibility to drive a specific mode into a non-linear regime while maintaining the linear response of the other mode. This is of interest for pump and probe experiments when there is strong coupling between modes.[21,22,33] At higher temperatures the resonator starts showing non-linear response only at higher drive voltages. This translates into an increased dynamic range of the device at room temperature.

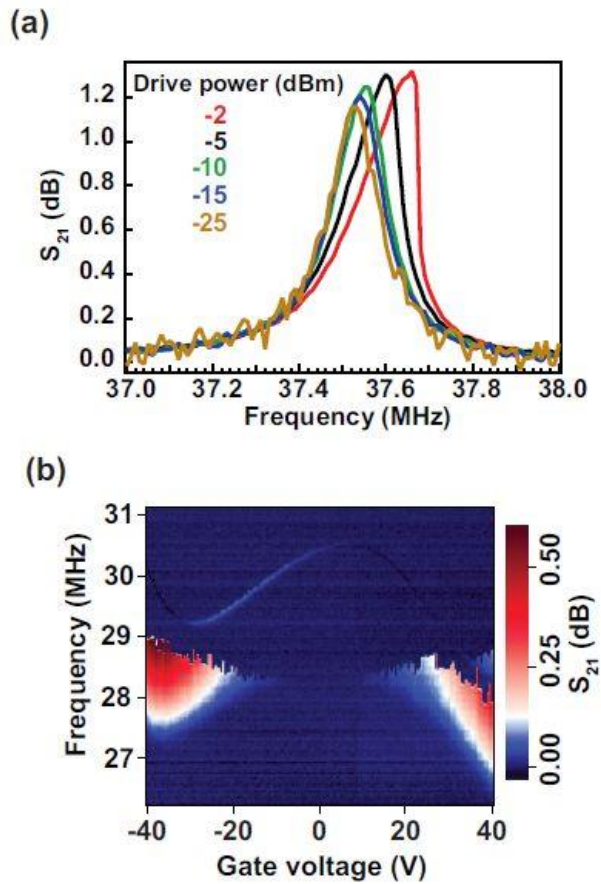

Figure 5: Measurement results from two devices. Device 1: a) Response of the resonator showing the transition to the nonlinear regime as the rf drive is increased from -25 dBm to -2 dBm at RT. Device 2: b) 2D plot of the non-linear response at RT.

In our devices even with low gate-resonator capacitive coupling (~ 10 aF) compared to previous work[10] (~ 40 aF) we measure signal (~ 0.1 dB above background) without taking advantage of the transconductance of the NW (fourth term in equation. 1). As the capacitive coupling in a CNT will be similar to that of a NW it may be possible to study CNT resonators on sapphire.

In conclusion, we have developed a simple fabrication technique for NEMS on sapphire substrate to enable large bandwidth actuation and detection. The measurements are performed at room temperature by direct electrical detection of the mechanical motion using a network analyzer. We observe variation of nonlinearity with temperature. This fabrication scheme can be applied to other NEMS (based on graphene and CNT) where the device size can be scaled down for studying non-linear mode coupling. In addition thermal properties of nanostructures may be studied.

Acknowledgement

We thank Akshay Naik for discussions. We acknowledge the Government of India and AOARD (Grant No. 124045) for support. JPM is supported by IBM student fellowship.

# Supporting information: Wide bandwidth nanowire electromechanics on insulating substrates at room temperature

T. S. Abhilash, John P. Mathew, Shamashis Sengupta, M. R. Gokhale, Arnab Bhattacharya, and Mandar M. Deshmukh

Department of Condensed Matter Physics and Materials Science, Tata Institute of Fundamental Research, Homi Bhabha Road, Mumbai 400005, India
E-mail: deshmukh@tifr.res.in

## Conductance measurement:

Fig. 1 shows the conductance variation with applied dc side gate voltage. This shows an n-type field effect transistor behavior.

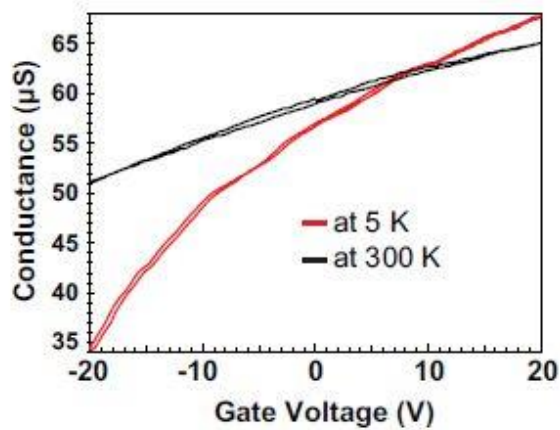

Figure 1: Conductance as a function of gate voltage.

## Resonance measurement

Room temperature measurements of mechanical resonance are carried out using a network analyzer[1–5] with $V_d^{dc}$, $V_g^{dc}$ and drive power as parameters. Measured resonance response as a function of $V_g^{dc}$ at a drain bias of $V_d^{dc}$ = - 100 mV is shown in Fig. 2. Both in-plane and out-of-plane modes are observed.

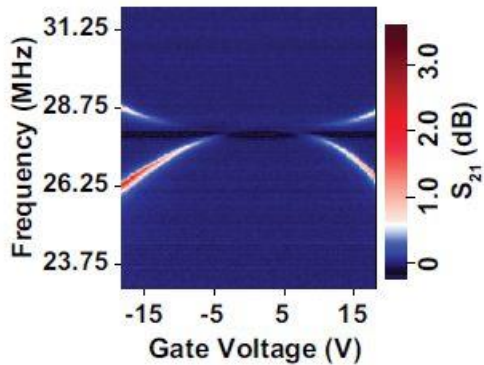

Figure 2: Shows color plot of $S_{21}$ as a function of driving frequency and $V_g^{dc}$ at room temperature.

Prior to each measurement the network analyzer is calibrated "through" the device keeping $V_g^{dc}$ = 0 and with an applied $V_d^{dc}$. Fig. 3 and inset shows the resonance response before and after background correction.

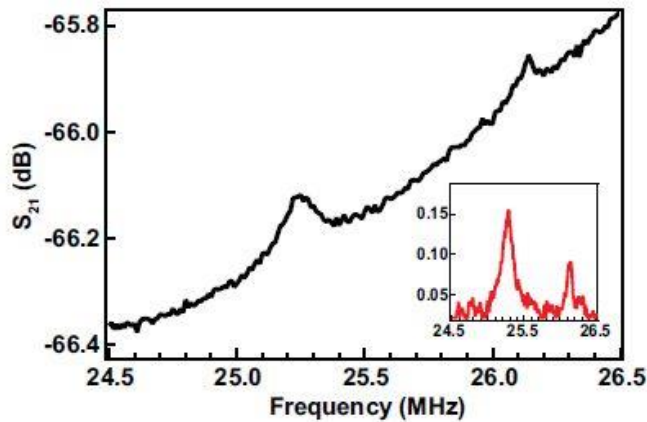

Figure 3: Response of the device before amplification and background correction. (Inset: after calibration).

## Oscillation amplitude

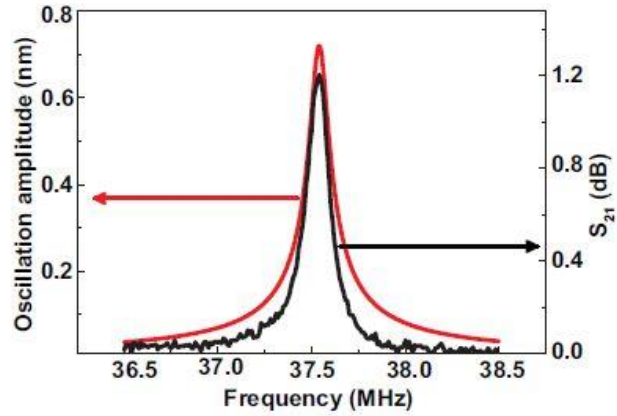

Figure 4: Calculated oscillation amplitude (red) and experimentally measured response (black).

We have calculated oscillation amplitude using the equation[3]

$$\tilde{z} = -\frac{1}{m}\frac{C_g}{z_0}V_g\tilde{V}_g\left(\frac{1}{\omega_0^2 - \omega^2 + j\omega_0\omega/Q}\right) \quad (1)$$

where, $m$ and $\omega_0$ is the mass and resonant frequency of nanowire resonator, $C_g$ and $z_0$ are the capacitance and distance between suspended nanowire and side gate, $\omega$ is the driving frequency and $Q$ the quality factor. The capacitance is $\sim$ 10 aF. $V_g$ = 20 V, $\tilde{V}_g$ = 20 mV and $z_0$ = 200 nm. The $S_{21}$ response and calculated value of amplitude at mechanical resonance, is shown in Fig. 4.

## Low temperature response

Fig. 5 shows line plots of measured $S_{21}$ as a function of driving frequency at 5K. This measurements are carried out with a drive power of -10 dBm. The resonance peak direction changes with positive and negative $V_d^{dc}$. At low temperatures, the resonant frequency of the devices increases. The resonance frequency shifts from 32.8 MHz to 39.84 MHz for the in-plane mode and 33.8 MHz to 40.82 MHz for the out-of-plane mode at 5K.

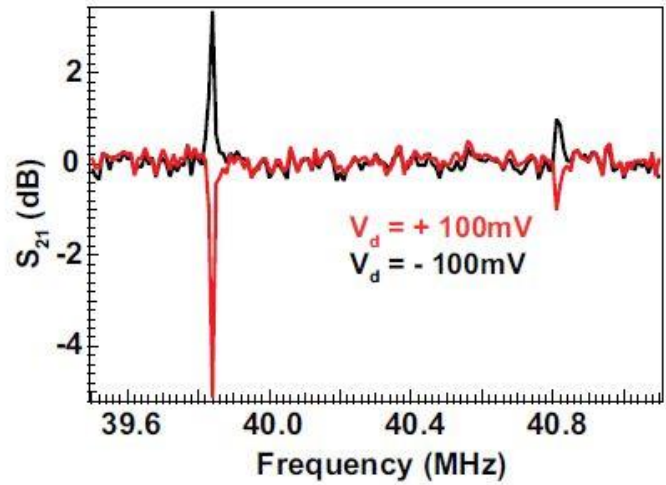

Figure 5: Line plots of $S_{21}$ as a function driving frequency at positive and negative $V_d^{dc}$ at 5K.

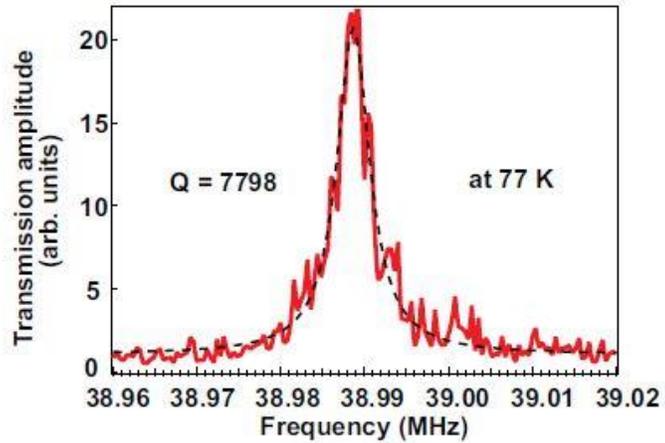

Figure 6: Shows Q of the device at 77K.

From room temperature down to 77 and 5K, quality factor (Q) increases by two orders of magnitude and becomes 7798 at 77K (Fig. 6) and 9958 at 5K. The resonance curves are fit to a Lorentzian to extract values Q. The fitted curve is shown in black line.

Fig. 7 shows a typical resonant response of the wire, where the vibration is in the plane of the side gate. As the drive power is increased, the response becomes nonlinear. At low temperature (5K) the onset of nonlinearity occurs at higher drives.

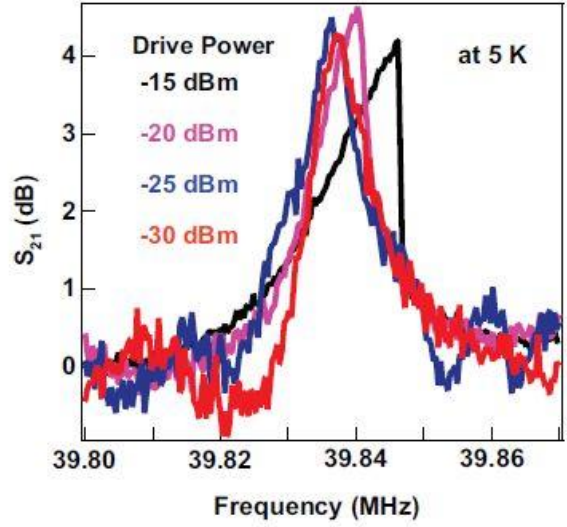

Figure 7: Nonlinear response at 5 K. Line attenuation is -15 dB.

## Temperature profile of the nanowire

The change in resonant frequency with source drain bias voltage (Fig. 3d) is attributed to thermal expansion of the nanowire when a current passes through it. Joule heating is used to calculate the temperature along the length of the NW with its ends kept at a fixed bath temperature $T_b$. The temperature as a function of the position, $x$, on a NW of length $L$, cross sectional area $A$, and thermal conductivity $\kappa$ with a current $I$ passing through it, is given by:

$$T(x) = \frac{R}{R'(e^{-\beta L} - e^{\beta L})}[(e^{-\beta L} - 1)e^{\beta x} + (1 - e^{\beta L})e^{-\beta x}] + T_b - \frac{R}{R'} \quad (2)$$

where $R$ is the resistance of the wire, $R' = \frac{dR}{dT}$, and $\beta = -\frac{I^2 R'}{L\kappa A}$. The temperature profile along the length of the NW is shown in Fig. 8. The heating is maximum at the center of the NW and reaches a temperature of $\sim 370$ K. Integrating the thermal expansion coefficient with the infinitesimal temperature along the length of the nanowire gives the increase in length to be $\sim 1$ nm. The corresponding strain can cause a resonant frequency shift[6] of a few MHz. To verify the change in resonant frequency due to thermal expansion, the resonant response of the NW was measured at 400 K. The resonant frequency is seen (Fig. 9) to shift by $\sim 1.5$ MHz.

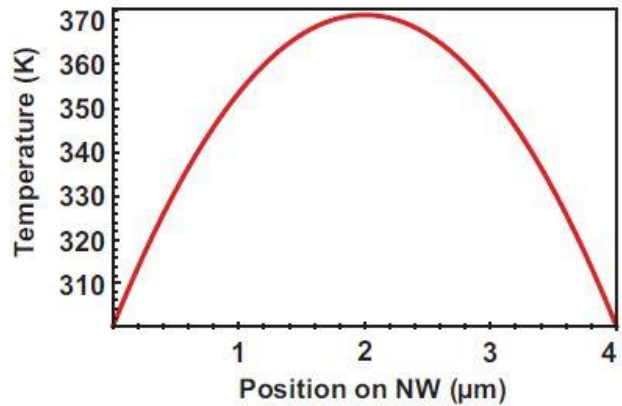

Figure 8: Temperature profile along the length of the NW.

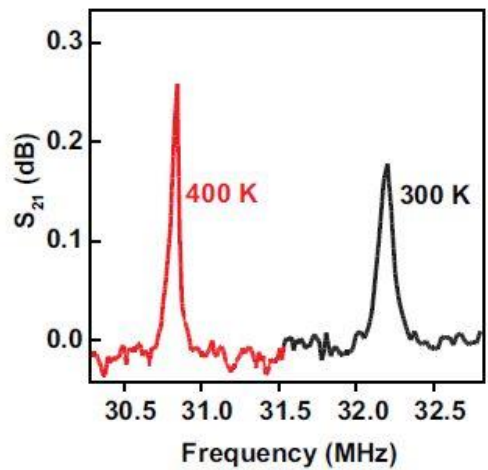

Figure 9: Response of the NW resonator at room temperature and 400 K. The resonant frequency is seen to decrease at 400 K.